\documentclass[aps,prx,showpacs,floatfix,twocolumn,superscriptaddress,longbibliography]{revtex4-2}
\usepackage{amsmath}
\usepackage{amssymb}
\usepackage{amstext}
\usepackage{amsopn}
\usepackage{amsfonts}
\usepackage{amsxtra}
\usepackage[english]{babel}
\usepackage{graphicx}
\usepackage{float}
\usepackage{subfigure}
\usepackage{bm}
\usepackage{multirow}
\usepackage{mathrsfs}
\usepackage{mathtools}
\usepackage{dcolumn}
\usepackage{upgreek}
\usepackage{ulem}

\usepackage[colorlinks,linkcolor=blue,urlcolor=blue,citecolor=blue]{hyperref}
\usepackage{color}
\usepackage{url}
\usepackage{breakurl}
\usepackage{relsize} 

\newcommand{\equa}[1]{Eq.~\eqref{#1}}
\newcommand{\unit}[1]{\,\mathrm{#1}} 
\newcommand{\fig}[1]{Fig.~\ref{#1}}

\begin{document}

\title{Revealing the frequency-dependent oscillations in nonlinear terahertz response induced by Josephson current}

\author{Sijie Zhang}
\email{sjzh@pku.edu.cn}
\affiliation{International Center for Quantum Materials, School of Physics, Peking University, Beijing 100871, China}
\author{Zhiyuan Sun}
\email{zysun@tsinghua.edu.cn}
\affiliation{State Key Laboratory of Low-Dimensional Quantum Physics and Department of Physics, Tsinghua University, Beijing 100084, China}
\author{Qiaomei Liu}
\author{Zixiao Wang}
\author{Qiong Wu}
\author{Li Yue}
\author{Shuxiang Xu}
\author{Tianchen Hu}
\author{Rongsheng Li}
\author{Xinyu Zhou}
\author{Jiayu Yuan}
\affiliation{International Center for Quantum Materials, School of Physics, Peking University, Beijing 100871, China}
\author{Genda Gu}
\affiliation{Condensed Matter Physics and Materials Science
Department, Brookhaven National Lab, Upton, New York 11973, USA}
\author{Tao Dong}
\author{Nanlin Wang}
\email{nlwang@pku.edu.cn}
\affiliation{International Center for Quantum Materials, School of Physics, Peking University, Beijing 100871, China}
\affiliation{Beijing Academy of Quantum Information Sciences, Beijing 100913, China}

%

%
\begin{abstract}
Nonlinear responses of superconductors to intense terahertz radiation has been an active research frontier. Using terahertz pump-terahertz probe spectroscopy, we investigate the $c$-axis nonlinear optical response of a high-temperature superconducting cuprate. After excitation by a single-cycle terahertz pump pulse, the reflectivity of the probe pulse oscillates as the pump-probe delay is varied. Interestingly, the oscillatory central frequency scales linearly with the probe frequency, a fact widely overlooked in pump-probe experiments. By theoretically solving the nonlinear optical reflection problem on the interface, we show our observation is well explained by the Josephson-type third-order nonlinear electrodynamics, together with the emission coefficient from inside the material into free space. The latter results in a strong enhancement of emitted signal whose physical frequency is around the Josephson plasma edge. Our result offers a benchmark for and new insights into strong-field terahertz spectroscopy of related quantum materials.
\end{abstract}

\pacs{\ }
\maketitle

%
%
The electrodynamic responses of high-temperature superconducting cuprates (HTSCs) are highly anisotropic. Conducting carriers are substantially constrained within the two-dimensional CuO$_2$ layers, while coherent  out-of-plane (c-axis) charge transport is not allowed in the normal state \cite{RN1241}. Below the transition temperature T$_{c}$, neighboring CuO$_2$ layers are coupled by the Josephson tunneling of Cooper pairs \cite{anderson1992experimental,PhysRevLett.68.2394,PhysRevLett.72.2263,Savel'ev_2010}. At the linear response level, the c-axis optical conductivity has a Drude form with a zero scattering rate and a plasma frequency being the Josephson plasmon mode (JPM) frequency $\omega_{\rm JPR}$. This means that weak electromagnetic fields with frequency below $\omega_{\rm JPR}$ are forbidden to propagate due to the screening of Josephson currents, while those above $\omega_{\rm JPR}$ are admissive, leading to a sharp Josephson plasmon edge (JPE) near $\omega_{\rm JPR}$ in the c-axis reflectance spectrum \cite{PhysRevLett.69.1455}. The appearance of the JPE implies the formation of three-dimensional superconductivity and the emergence of JPM along the $c$-axis.

\begin{figure}[t]
\centering
\includegraphics[width=8cm, trim=0 0 0 0,clip]{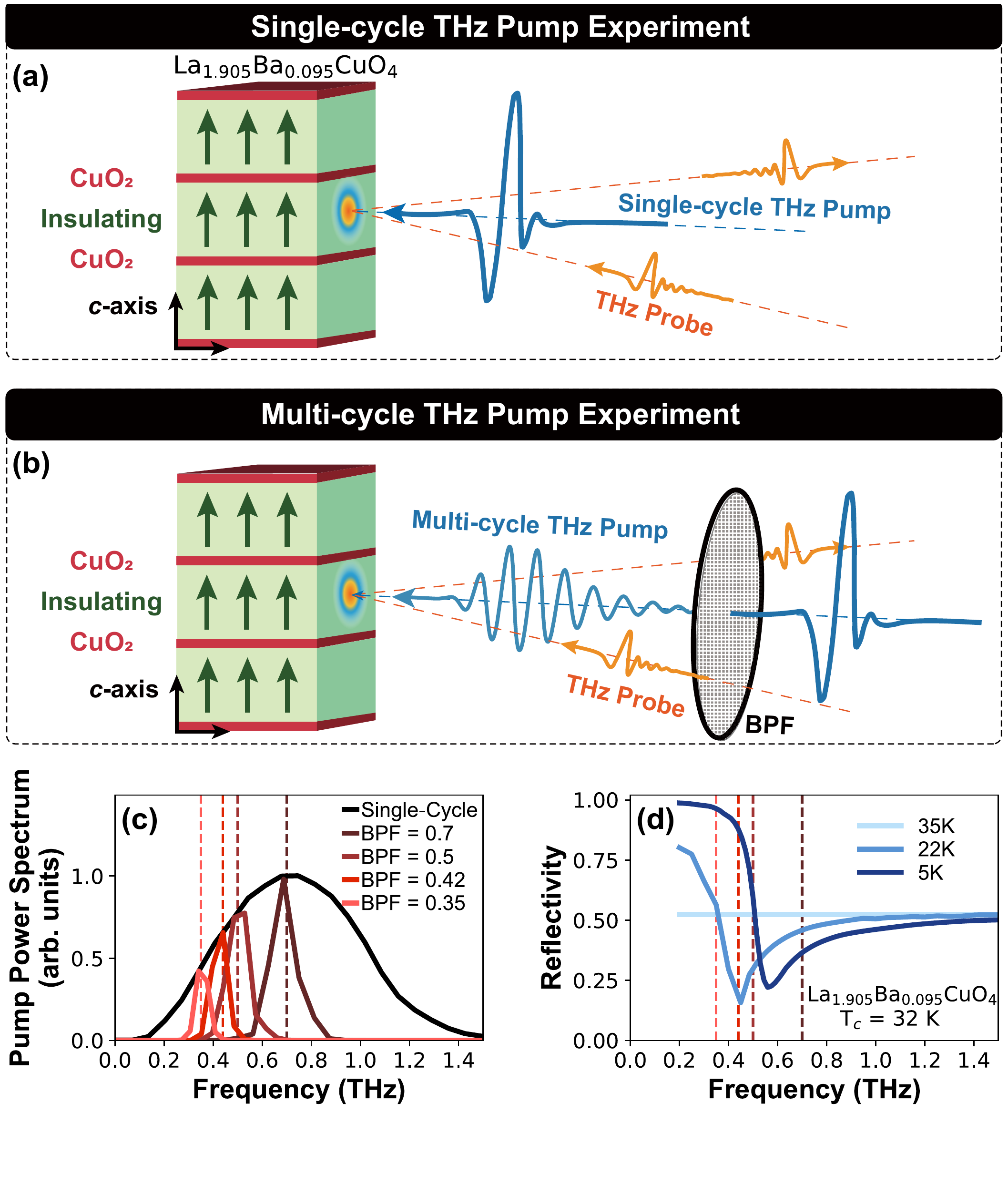}\\
\caption{Experimental setup and sample characterization. (a) and (b) Schematics of the THz pump-THz probe spectrometers with broad- and narrow-band THz pump. Both THz pump and probe are paralleled with $c$-axis of La$_{1.905}$Ba$_{0.095}$CuO$_4$. Green arrows indicate the direction of the Josephson tunneling current. BPF represents a band-pass filter. (c) Power spectra of the broad-band THz pump (black) and narrow-band ones with center frequencies of 0.35, 0.42, 0.5, and 0.7 THz (colored). (d) C-axis reflectance spectra where sharp Josephson plasmon edges show up at T $<$ T$_c$.}
\label{Fig:1}
\end{figure}

In the nonlinear regime, the nonlinear responses induced by Josephson tunneling may also be cast in the form of odd-order nonlinear susceptibilities (see Supplementary information \cite{supp}). Since the JPE of HTSCs usually lies in the terahertz (THz) range, the nonlinear response induced by intense THz radiations are anticipated to exhibit special features. Thanks to the advent of femtosecond laser techniques, strong-field THz pulse with stable carrier-envelope phase (CEP) has been a powerful tool for detecting and manipulating exotic quantum states, by introducing nonlinear processes without injecting overwhelmed energy \cite{Bi2Se3_floqet_1,topo_light_induced_hall,RN391,mode_selective_magnetic_NiO,electromagnon_THz,TmFeO3_THz_orbit_excitation,SrTiO3_drive_LiNbO3,mode_selective_SrTiO3_THz,MoTe2_UED,RN398}.
Recently, several nonlinear electromagnetic responses related to JPM using CEP-stable strong-field THz pulses were reported \cite{rajasekaran2016parametric,rajasekaran2018probing,LSCO_THG,YBCO_absence}. Rajasekaran \textit{et al.} found that, in the THz pump-THz probe measurement  along the $c$-axis of the superconducting La$_{1.905}$Ba$_{0.095}$CuO$_4$, temporal oscillations centering at 2$\omega_{\rm JPR}$ appear in the pump-probe delay process after the excitation of single-cycle THz pulses~\cite{rajasekaran2016parametric}. Subsequently, they also observed a giant third-harmonic generation (THG) of single-cycle broad-band THz pulses \cite{rajasekaran2018probing}.

In this paper, we performed THz pump-THz probe spectroscopy on La$_{1.905}$Ba$_{0.095}$CuO$_4$ and investigated the out-of-plane transient optical responses after the excitations of single-cycle THz pulses and multi-cycle ones. We observed long-lasting temporal oscillatory signals in the pump-probe decay process of the superconducting state after THz excitations. Surprisingly, by discerning the oscillatory behavior at different THz probe frequencies, we found that the frequency of oscillations induced by the single-cycle THz pump depends linearly on the probe frequency.  In contrast, multi-cycle THz pump centering at $\omega_{\rm pump}$ induces single-frequency oscillations near 2$\omega_{\rm pump}$, whose amplitude arrives at a maximum as $\omega_{\rm pump}$ approaches $\omega_{\rm JPR}$. By plotting the nonlinear signal on the two-dimensional plane of the pump and probe frequencies, we reveal the origin of these behaviors as an enhancement when the physical frequency of the signal is around $\omega_{\rm JPR}$. This is explained simply by the emission coefficient that we derived from solving the nonlinear electromagnetic (EM) problem for the interface, while the resonant excitation of a single collective mode is not assumed. Our result shows that in pump-probe nonlinear spectroscopy, this emission coefficient needs to be included in order to extract meaningful information of the probed material.

\section{Results}

\begin{figure*}[t]
\centering
\begin{minipage}{5.9cm}
\centering
\includegraphics[width=1\textwidth]{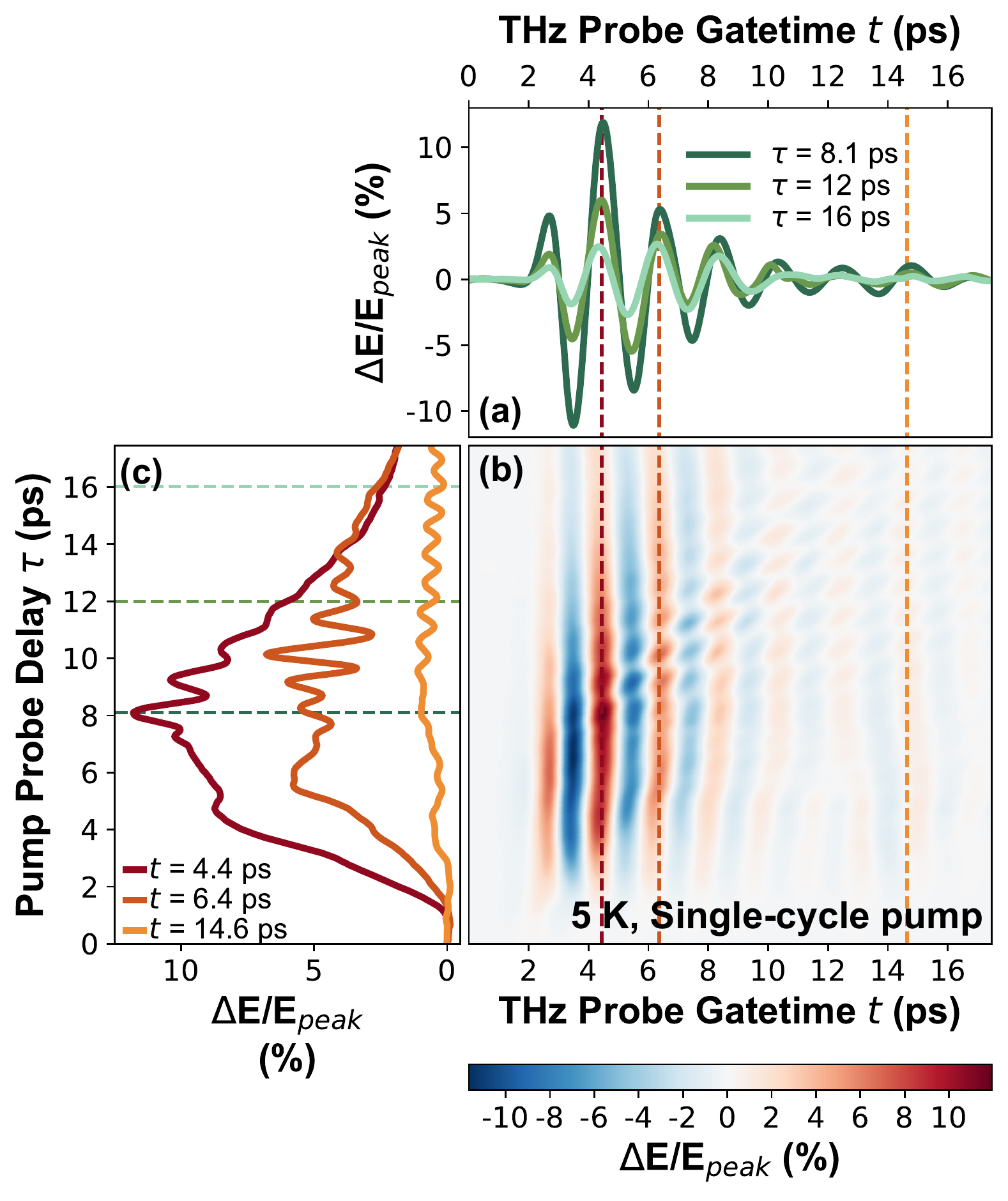}
\end{minipage}
\hfill
\begin{minipage}{6.1cm}
\centering
\includegraphics[width=1\textwidth]{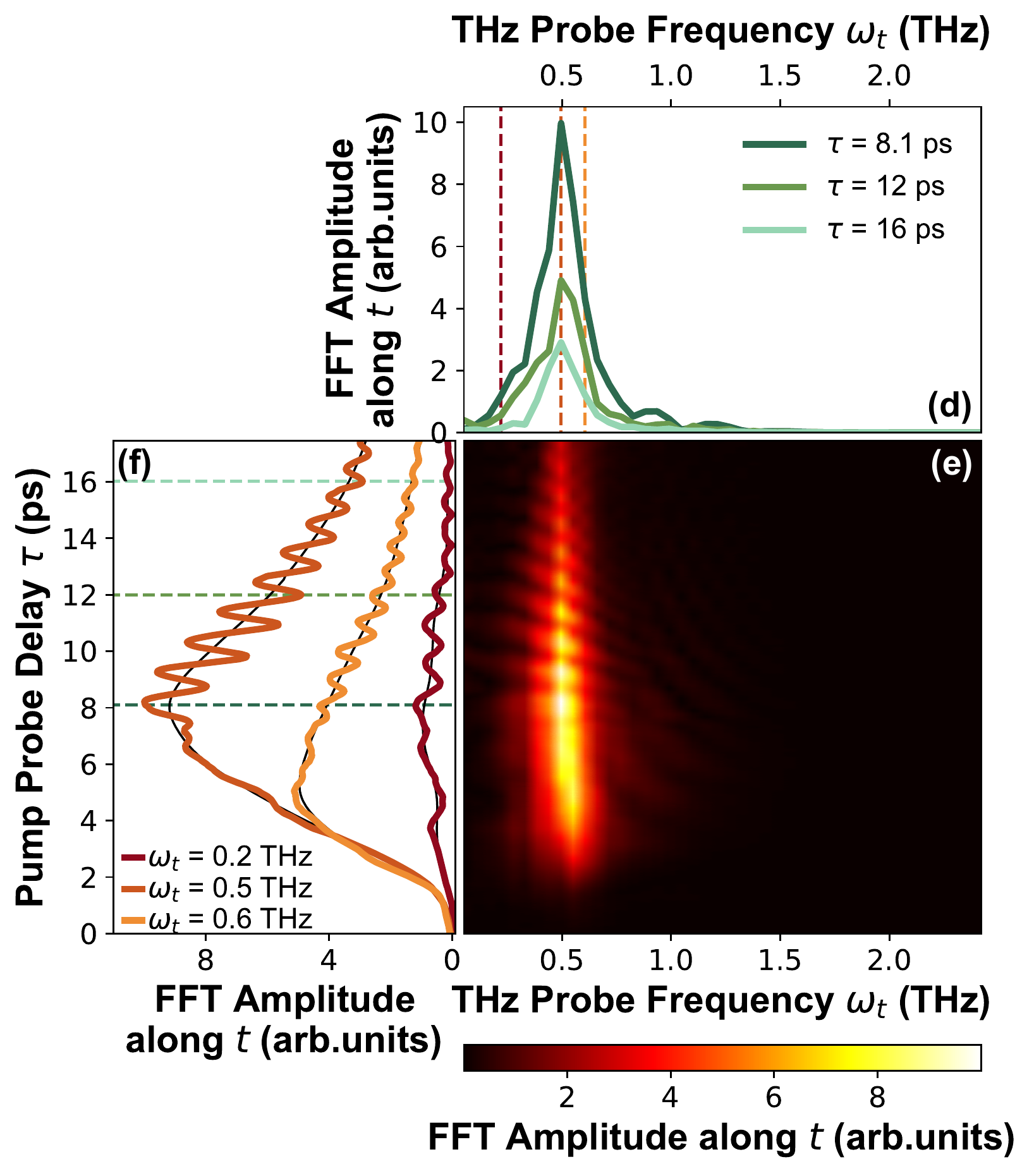}
\end{minipage}
\hfill
\begin{minipage}{4.6cm}
\centering
\includegraphics[width=1\textwidth]{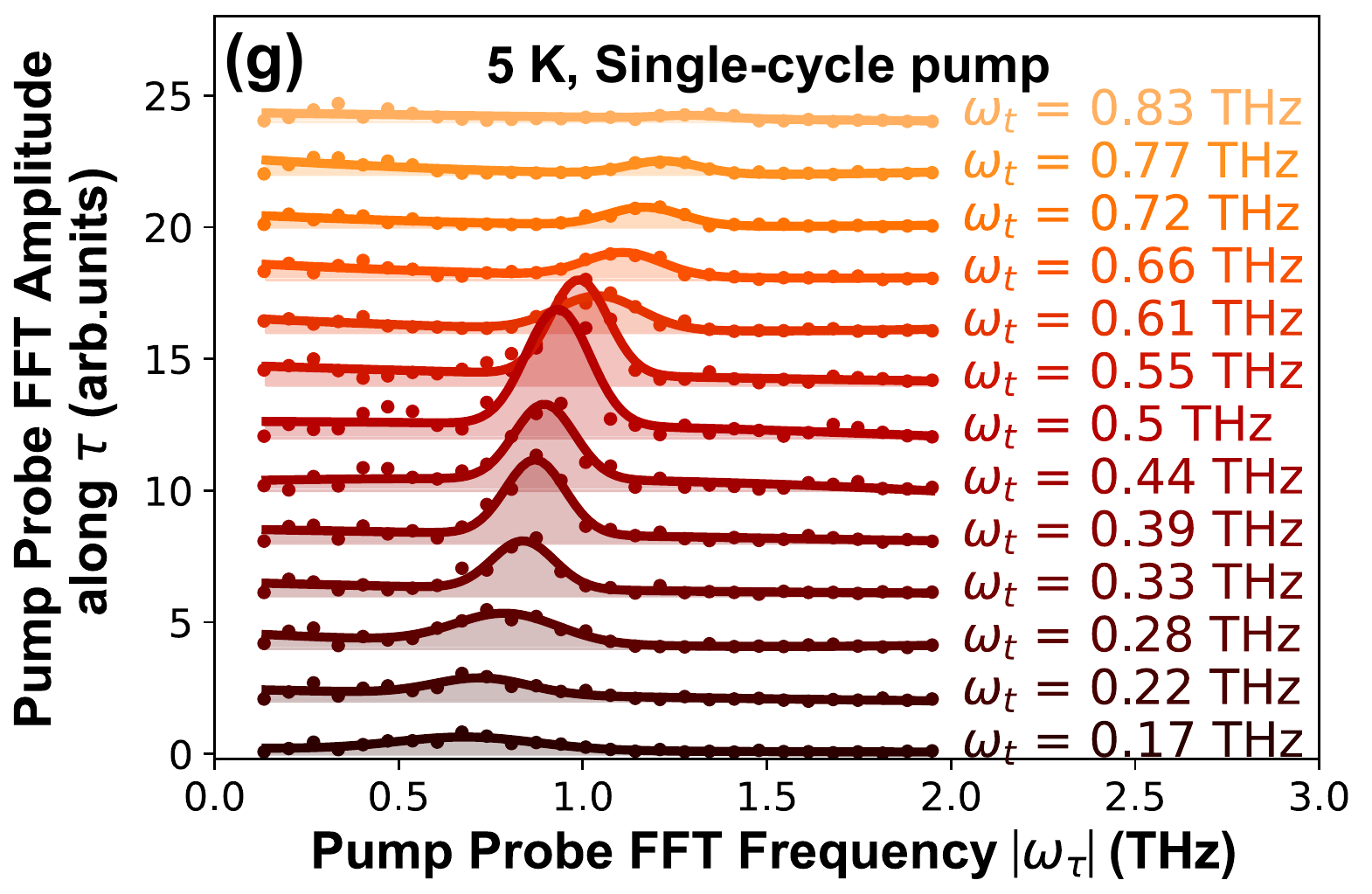}
\includegraphics[width=1\textwidth]{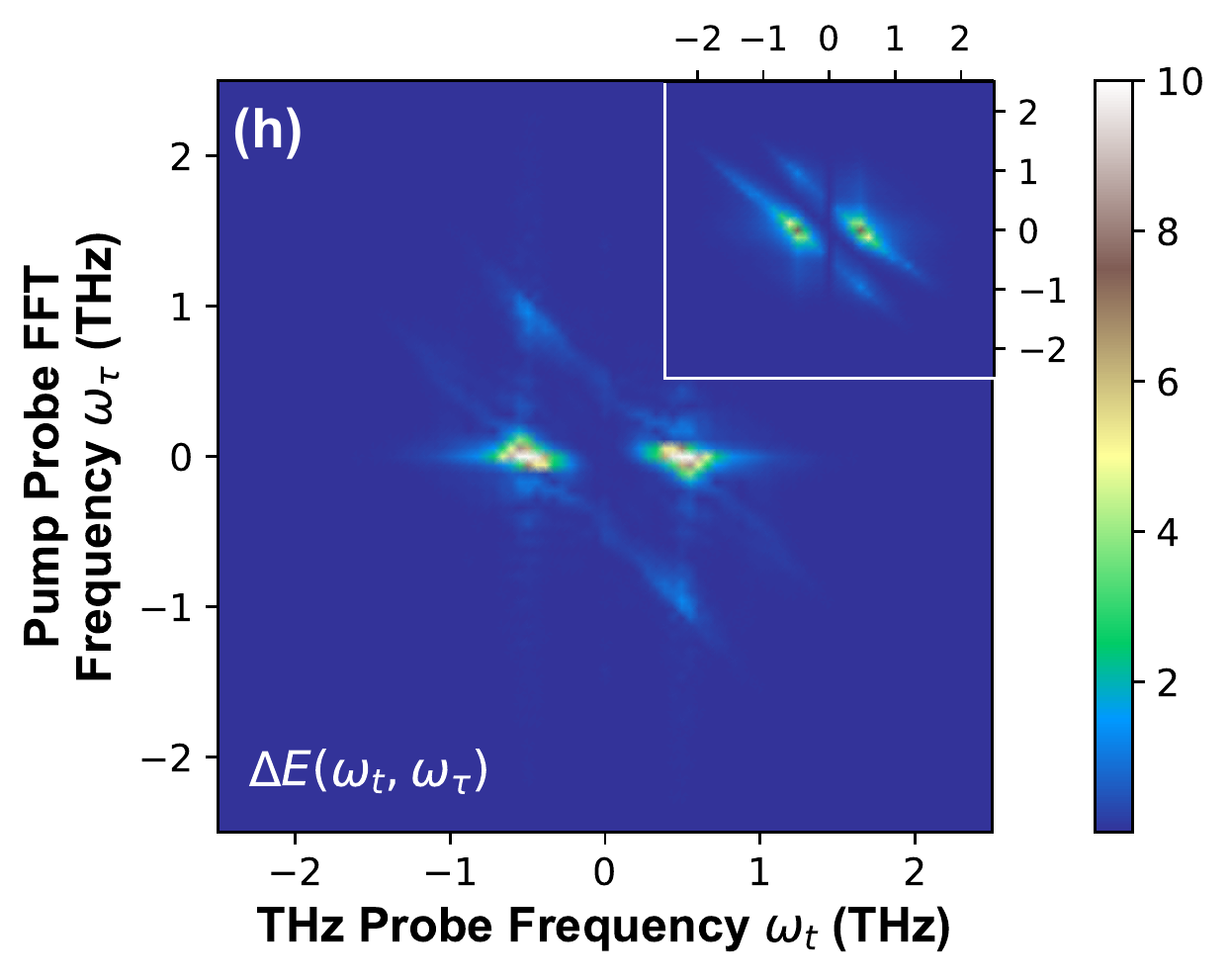}
\end{minipage}
\setlength{\abovecaptionskip}{0.1cm}
  \caption{Single-cycle THz pump experiment at 5 K. (a) The pump-induced changes along electro-optic sampling gatetime $t$ of reflected THz probe pulse at certain pump-probe delay time $\tau$, $\Delta E(t, \tau)$. (b) A two-dimensional time-domain $\Delta E(t, \tau)$ color plot. (c) The decay profiles of $\Delta E(t, \tau)$ at three gatetime, in which clear oscillations are observed. The oscillation period approaches 1 ps as increasing $t$ from 4.4 to 14.6 ps. (d) The pump-induced changes in the frequency domain $\omega_t$, $\Delta E(\omega_t, \tau)$. (e) The two-dimensional color plot of $\Delta E(\omega_t, \tau)$ presents a temporal modulation in a series of backward-slashes pattern along $\tau$. (f) The decay profiles of $\Delta E(\omega_t, \tau)$ at three representative $\omega_t$. The oscillating signals can be obtained by subtracting away the non-oscillating components plotted as the thin black curves. (g) FFT amplitude of the extracted oscillatory signals in $\Delta E(\omega_t, \tau)$ is plotted as scatters and fitted with solid curves. (h) The two-dimensional FFT of the measured $\Delta E(t, \tau)$ shown in (b), $\Delta E(\omega_t, \omega_{\tau})$. Inset: $\Delta E(\omega_t, \omega_{\tau})$ estimated by numerically solving the third-order nonlinear responses.}
\label{Fig:2}
\end{figure*}

Figure \ref{Fig:1} summarizes the configuration of the THz pump-THz probe experiments. The schematic setups for single- and multi-cycle THz pump experiments are illustrated in Fig. \ref{Fig:1} (a) and (b). The single-cycle THz pump is centered at 0.7 THz with peak electric field of $\sim$ 200 kV/cm, shown as the black curve in Fig. \ref{Fig:1} (c). The multi-cycle THz pump pulses are centered at 0.35, 0.42, 0.5, and 0.7, with peak electric fields of 30, 40, 60, and 70 kV/cm, respectively. The polarization of the THz pump and probe is set to be parallel to the $c$-axis of the HTSC sample. For the single-layer HTSC La$_{2-x}$Ba$_{x}$CuO$_4$, there is only a uniform Josephson coupling strength along the $c$-axis. In addition, the stripe order within CuO$_2$ layers, which can restrain the Josephson tunneling and compete with superconductivity, is nearly absent when the doping level is away from $x$ = 0.125 \cite{PhysRevB.85.134510}. To avoid the complicated interactions between different JPMs and those between JPM and stripe order, we measured the sample without a stripe order phase, La$_{1.905}$Ba$_{0.095}$CuO$_4$, whose T$_c$ is 32 K. The $c$-axis reflectance spectra in the THz regime is presented in Fig. \ref{Fig:1} (d). In the normal state, an insulating-like response is observed. Once entering the superconducting state, a sharp JPE shows up. As temperature decreases, JPE shifts to higher energy due to the increase of superfluid density.

Figure \ref{Fig:2} summarizes the out-of-plane transient responses at 5 K after the excitation by single-cycle THz pulses. The raw data is the pump-induced changes of reflected THz probe scanned along EOS gatetime $t$ (defined as the time difference between the gatetime and the arrival time of the probe pulse) at a certain pump-probe delay time $\tau$ (the time difference between the gatetime and the arrival time of the pump pulse), \textit{i.e.} $\Delta E(t, \tau)$ shown in Fig. \ref{Fig:2} (a). A two-dimensional time-domain color plot of $\Delta E(t, \tau)$ shown in Fig. \ref{Fig:2} (b) is achieved by scanning over $\tau$. The decay profiles of $\Delta E(t, \tau)$ at representative $t$ shown in Fig. \ref{Fig:2} (c) are obtained by cutting along $\tau$-axis. To further clarify the transient responses, the pump-induced changes in frequency domain are studied. By performing fast-Fourier transformation (FFT) on $\Delta E(t, \tau)$ along $t$-axis, $\Delta E(\omega_t, \tau)$ is determined as shown in Fig. \ref{Fig:2} (d). The pump-induced change primarily happens near $\omega_{\rm JPR}\sim$0.5 THz. The corresponding transient optical properties show a splitting of JPM induced by the THz pump as presented in Supplementary information (SI) \cite{supp}, similar to the responses after mid- and near-infrared excitations \cite{PhysRevB.98.020506,PhysRevB.98.224507}. Figure \ref{Fig:2} (e) shows the two-dimensional color plot of $\Delta E(\omega_t, \tau)$, which is temporally modulated by a series of backward-slash patterns along $\tau$. The slopes of those slashes are different, which suggests the $\tau$-axis oscillation period is not a constant along THz probe frequency $\omega_{t}$. Figure \ref{Fig:2} (f) shows the pump-probe decay profiles at three representative $\omega_t$.

Those oscillatory signals along $\tau$ are extracted by subtracting away the non-oscillating pump-probe curves (thin black curves in Fig. \ref{Fig:2} (f)). The oscillatory amplitude is shown in Fig. \ref{Fig:2} (g). There is a clear blueshift of the oscillatory center frequency as increasing the THz probe frequency $\omega_t$, which indicates a linear relationship between $\omega_{\tau}$ and $\omega_t$. Identical oscillatory behaviors also exist in the transient optical constants, as presented in SI. Although temporal oscillations in pump-probe spectral experiments have been widely observed, those oscillations usually center at a fixed frequency and correspond to an excitation of some specific collective mode \cite{PhysRevB.100.165131} such as coherent phonons, amplitude modes in charge density wave compounds, and Higgs modes. The linear varying behavior reported here has not been reported before. For the THz-induced transient out-of-plane responses of La$_{1.905}$Ba$_{0.095}$CuO$_4$ reported in Ref. \cite{rajasekaran2016parametric}, the oscillation along the pump-probe delay $\tau$ centers at the fixed frequency 2$\omega_{\rm JPR}$ since they were looking at the single at a fixed $t$, which may be viewed as a special case of the linear dependent relation reported here.

Fig. \ref{Fig:2} (h) shows the two-dimensional FFT of the measured $\Delta E(t, \tau)$ shown in Fig. \ref{Fig:2} (b), $\Delta E(\omega_t, \omega_{\tau})$. There are four bright spots in Fig. \ref{Fig:2} (h), which are located at ($\pm$0.5, 0), (0.5, $-$1) and ($-$0.5, 1). The bright spots at ($\pm$0.5, 0) correspond to the non-oscillating component along the $\tau$-axis in Fig. \ref{Fig:2} (f), and the ones at (0.5, $-$1) and ($-$0.5, 1) correspond to the oscillating component. It is noteworthy that the bright spots at (0.5, $-$1) and ($-$0.5, 1) are stretched along the direction $\omega_t + \omega_\tau = \pm0.5$, which reflects the linear relationship between $\omega_{\tau}$ and $\omega_t$ observed in Fig. \ref{Fig:2} (g).

\begin{figure*}[t]
\centering
\begin{minipage}{6.5cm}
\centering
\includegraphics[width=1\textwidth]{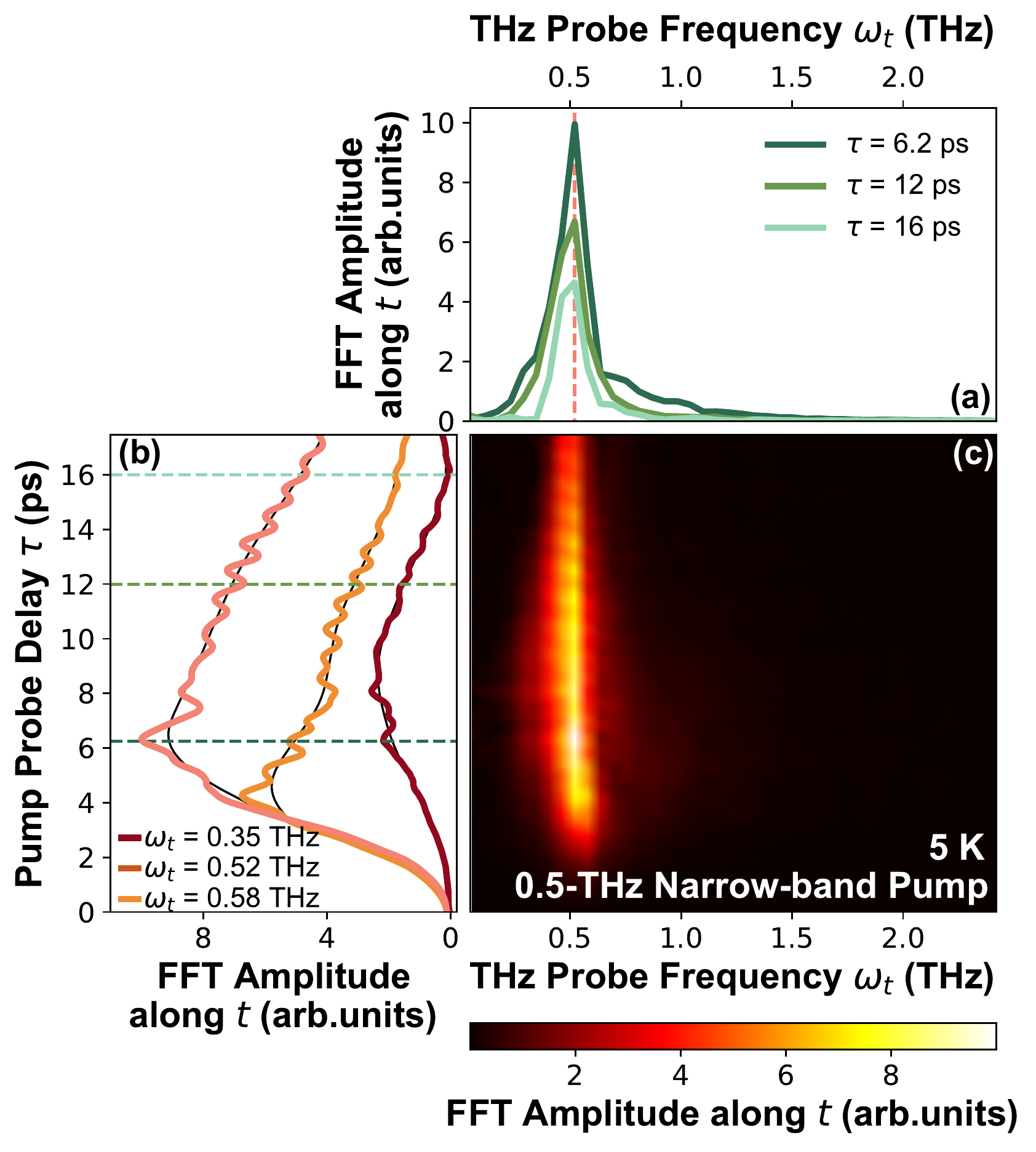}
\end{minipage}
\hfill
\begin{minipage}{5cm}
\centering
\includegraphics[width=1\textwidth]{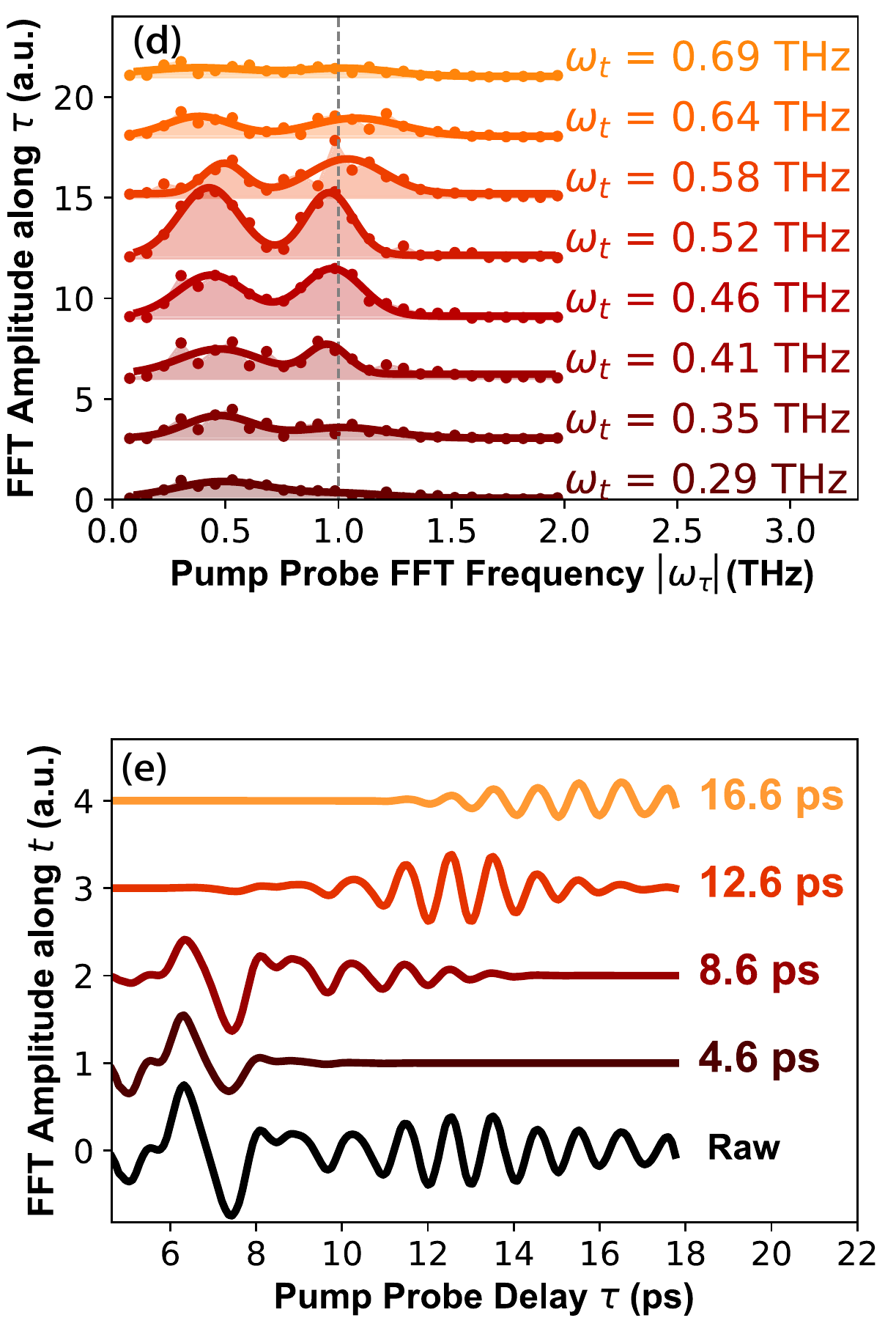}
\end{minipage}
\hfill
\begin{minipage}{5cm}
\centering
\includegraphics[width=1\textwidth]{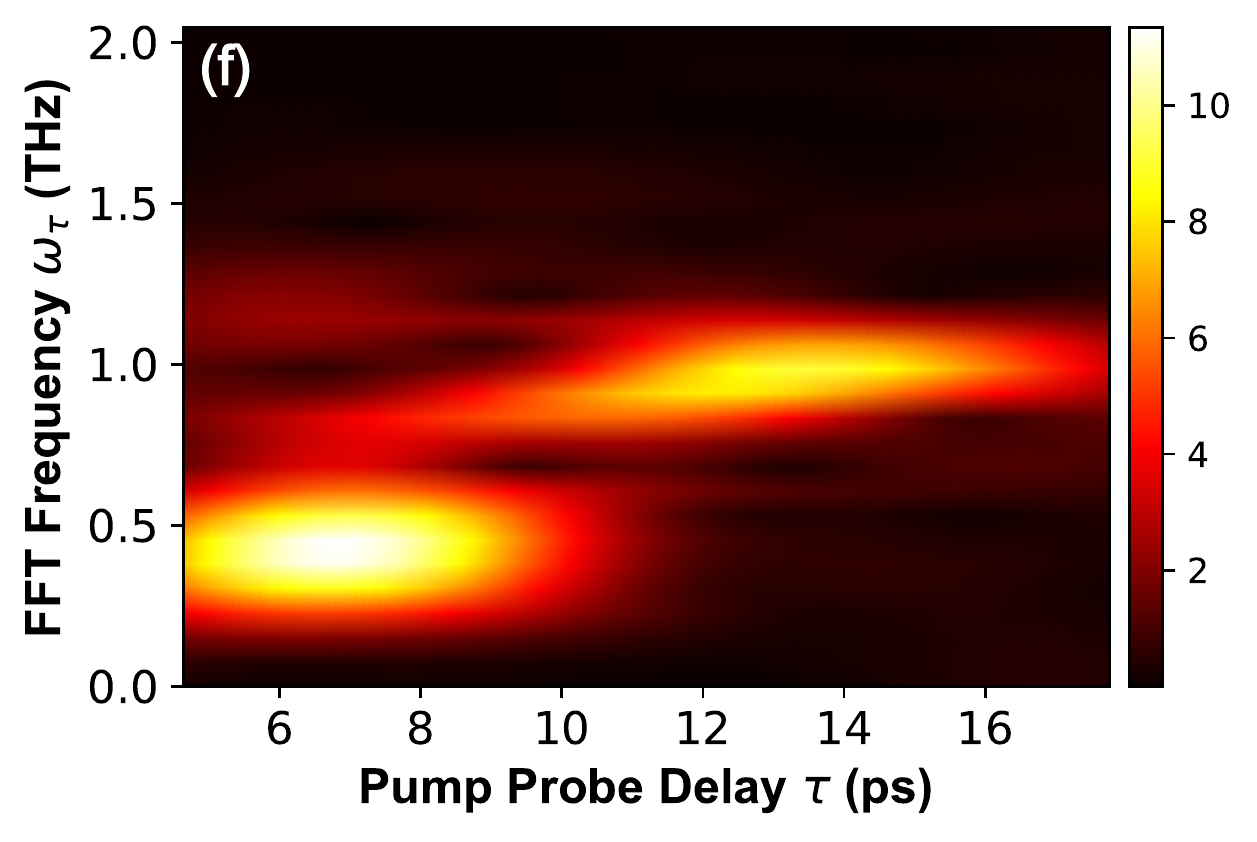}
\includegraphics[width=1\textwidth]{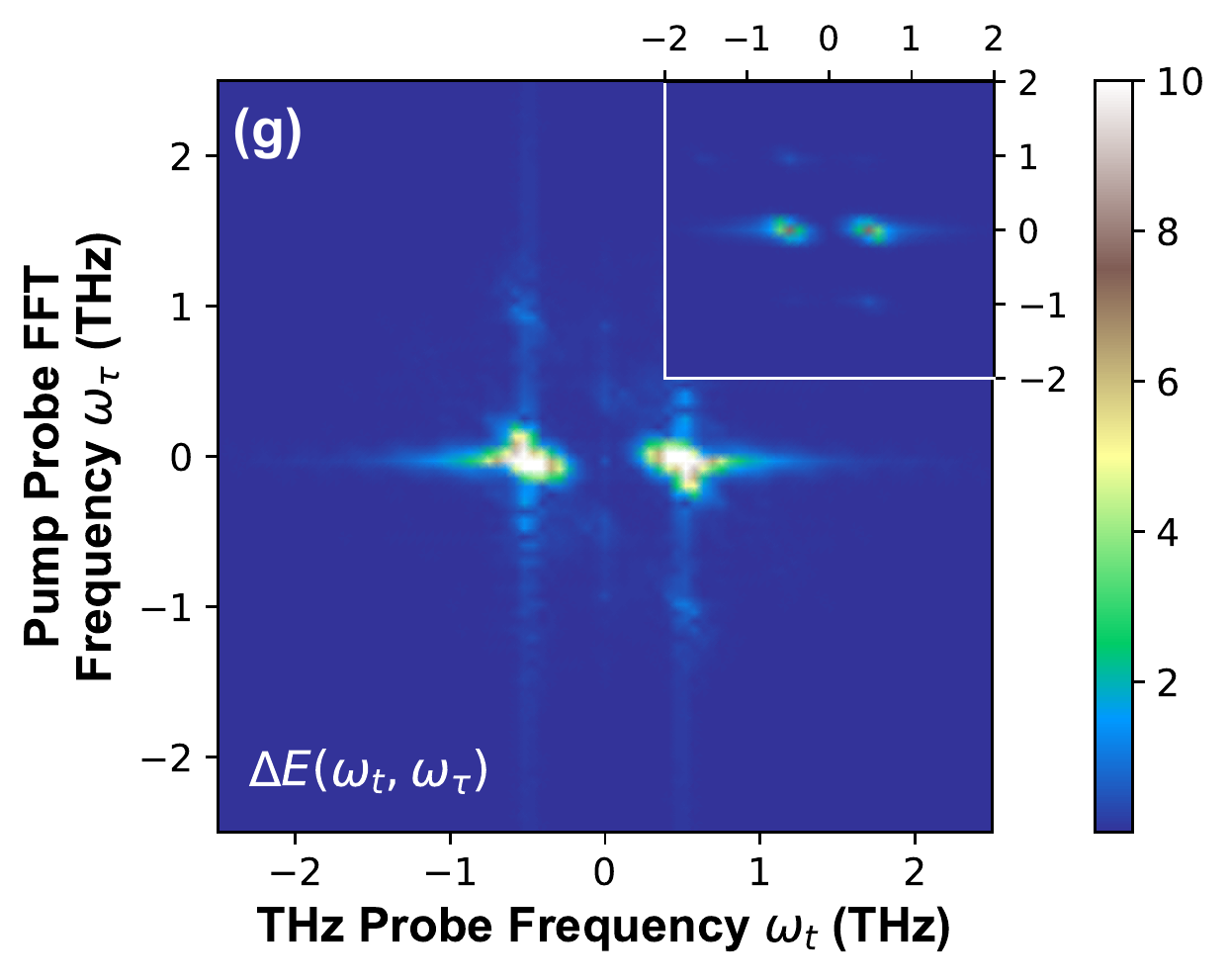}
\end{minipage}
\caption{Multi-cycle THz pump experiment at 5 K. (a) The pump-induced changes in the frequency domain $\omega_t$, $\Delta E(\omega_t, \tau)$, at three representative pump-probe delay time $\tau$ . (b) The decay profiles of $\Delta E(\omega_t, \tau)$ at three $\omega_t$, in which oscillatory signals can be obtained by subtracting the non-oscillating background plotted as thin black curves. (c) The two-dimensional color plot of $\Delta E(\omega_t, \tau)$, in which a temporal modulation is observed along $\tau$. (d) FFT amplitude of the extracted oscillatory signals in $\Delta E(\omega_t, \tau)$, which peaks near 0.5 and 1 THz. The grey dashed line is a guide to the eyes. (e) Frequency-resolved optical gating technique is used for analyzing the time-varying oscillation period in (b), by applying a gating window to the raw oscillation at $\omega_t$ = 0.52 THz (black curve). By moving the gating window continuously, dominating signals near $\tau$ are extracted (colored curves). (f) FFT amplitude of the windowed oscillations. The oscillations near $\omega_{\rm pump}$ dominate initial $\tau$ and long-lasting oscillations near 1 THz dominate the subsequent $t$. (g) The two-dimensional FFT of the measured $\Delta E(t, \tau)$, $\Delta E(\omega_t, \omega_{\tau})$, in which four bright spots can be observed. Inset: $\Delta E(\omega_t, \omega_{\tau})$ estimated by numerically solving $\equa{eqn:chi3}$.}\label{Fig:3}
\end{figure*}

We now try to interpret that linear dependence by THz nonlinear electrodynamics. Since there is  inversion symmetry in La$_{1.905}$Ba$_{0.095}$CuO$_4$, the leading nonlinear optical process must be due to the third-order nonlinear optical response. Indeed, the measured signal scales linearly with the probe field and  quadratically with the pump field. By solving the full EM problem, we derive the pump-induced change of the reflected probe field $\Delta E(t,\tau)$  as:
\begin{align}
\Delta E(t,\tau)=&
\sum_{\omega_1, \omega_2, \omega_3}F(\omega, \mathbf{k})\chi^{(3)}{\left(\omega_1, \omega_2,\omega_3\right)}
\notag\\
&
E_{\text {pump }}\left(\omega_1\right)E_{\text{pump}}\left(\omega_2\right)E_{\text {probe }}\left(\omega_3\right)
\notag\\
&
e^{-i\left(\omega_1+\omega_2\right) \tau-i \omega_3 t}
\label{eqn:chi3_time}
\end{align}
where $ E_{\text{pump}}$ and $E_{\text{probe}}$ are the electric fields of transmitted pump and probe pulses inside the sample, $\chi^{(3)}$ is the third order nonlinear susceptibility defined as the ratio of the third order polarization to the incident fields, $F$ is the emission coefficient of the third order polarization into far field, and $(\omega=\omega_1+\omega_2+\omega_3, \mathbf{k})$ is the frequency and momentum of the polarization generated from the third order nonlinear effect (See Section 3 of SI for derivation).
The Josephson relation predicts that
 $\chi^{(3)}(\omega_1, \omega_2, \omega_3) = \frac{(2e)^3}{3 !}J_c d^3 /[(\omega_1+ \omega_2+ \omega_3)\omega_1 \omega_2 \omega_3]$ where $J_c$ is the Josephson critical current density, $d$ is the inter-CuO$_2$ layer distance and $e$ is the elementary charge.
The Fourier transform of $\Delta E(t,\tau)$ is therefore:

\begin{align}
&\Delta E (\omega_t,\omega_\tau) 
\notag\\
=
&\sum_{\omega}
F(\omega_t+\omega_\tau,\mathbf{k}) \chi^{(3)}(\omega,\omega_\tau-\omega, \omega_t)
\notag\\
&E_{\text{pump}}(\omega) E_{\text{pump}}(\omega_\tau-\omega)
E_{\text{probe}}(\omega_t)
\,.
\label{eqn:chi3}
\end{align}
Since the signal is a product of two pumps and one probe, its frequency is equal to the sum of the frequency of each constitute. Thus $\Delta E (\omega_t,\omega_\tau)$ in \equa{eqn:chi3} means the amplitude of the third order signal with physical frequency $\omega_{\text{physical}}=\omega_\tau+\omega_t$. The incident pump/probe pulse are modeled as Gaussian functions reshaped by the transmission coefficient $T(\omega)$:
\begin{align}
 E_{\text{probe}}(\omega) =
 &T(\omega)
 E_{\text{pr}} (e^{-(\omega-\omega_{\text{pr}})^2/W^2_{\text{pr}}}
  \notag\\&+ e^{-(\omega+\omega_{\text{pr}})^2/W^2_{\text{pr}}} )
\,, \notag\\
  E_{\text{pump}}(\omega) =
  &T(\omega) E_{\text{pu}}  (e^{-(\omega-\omega_{\text{{pu }}})^2/W^2_{\text{pu}}} \notag\\& + e^{-(\omega+\omega_{\text{{pu }}})^2/W^2_{\text{pu}}} )
 \,.
 \label{eqn:profiles}
\end{align}
where  the central frequency $\omega_{\rm pu}$/$\omega_{\rm pr}$ and spectra widths $W_{\rm pu}$/$W_{\rm pr}$ are determined by measurements.
Note that $T(\omega)$ is calculated from the optical conductivity inferred by the reflectance spectrum (Fig. \ref{Fig:1} (d)).
Plugging \equa{eqn:profiles} into \equa{eqn:chi3} gives the spectrum of $\Delta E (\omega_t,\omega_\tau)$.

The inset of Fig. \ref{Fig:2} (h) presents the numerical result of \equa{eqn:chi3}, with $\omega_{\rm pu} = 0.7$ THz, $\omega_{\rm pr} = 1$ THz, $W_{\rm pu} = 0.6$ THz, and $W_{\rm pr} = 1$ THz. The result explains all the bright spots in Fig. \ref{Fig:2} (h).
Notably, the sharp bright lines along $\omega_t+\omega_\tau=\pm \omega_{\text{JPR}}$ may be explained by the sharp peaks in the emission coefficient $F(\omega, \mathbf{k})$ when the physical frequency $\omega=\omega_t+\omega_\tau$ of the third order nonlinear signal  matches $\pm \omega_{\text{JPR}}$, so that the wave vector {$\lvert$}$k_{\downarrow}${$\lvert$}={$\lvert$}$\sqrt{\epsilon_c \omega^2/c^2}${$\lvert$} of the emitted transverse EM mode  inside the sample vanishes. The signal at this frequency emits most easily out of the sample. The same coefficient also suppresses the signals in the regions $\omega_t, \omega_\tau>0$ and $\omega_t, \omega_\tau<0$. Therefore, the above nonlinear response is well explained by the third-order nonlinear electrodynamics combined with the JPE. We remark that our approach goes beyond solving the dynamics of a nonlinear harmonic oscillator by considering the full EM problem on the interface between the material and space. Similar bright spots may occur in the two-dimensional spectroscopy of few level systems \cite{PhysRevA.102.043514}, which merits further investigations.

We now investigate the out-of-plane transient responses to CEP-stable multi-cycle THz pulses, which has not been explored before. Figure \ref{Fig:3} presents the experimental results after 0.5-THz multi-cycle THz excitations. $\Delta E(\omega_t, \tau)$ also peaks around $\omega_{\rm JPR}$, with a temporal modulation along $\tau$, as shown in Fig. \ref{Fig:3} (a) and (b). Figure \ref{Fig:3} (c) shows the two-dimensional color plots of $\Delta E(\omega_t, \tau)$, in which nearly no $\omega_t$-dependent modulation is observed. Figure \ref{Fig:3} (d) shows the oscillation amplitude at different $\omega_t$, which all peak around $0.5 \unit{THz}$ and $1 \unit{THz}$. We note that those pump-induced oscillations are invisible after excitations of CEP-unstable narrow-band THz pulses radiated from a free-electron laser \cite{dienst2013optical}. Furthermore, the oscillation frequencies are nearly independent of THz probe frequency $\omega_t$ anymore. We also note that the oscillation period in Fig. \ref{Fig:3} (b) varies in time, which can be analyzed by a frequency-resolved optical gating technique \cite{PhysRevB.105.L100508}. As shown in Fig. \ref{Fig:3} (e), gating windows in a Gaussian form with a duration of 3 ps are applied to the extracted raw oscillatory signals. By moving the gating window continuously in the pump-probe delay, dominant signals near $\tau$ are extracted, of which the FFT amplitudes are presented in Fig. \ref{Fig:3} (f). That time-frequency distribution analysis shows that the quasi-single-cycle oscillation near 0.5 THz only exists during the initial $\tau$ while long-lasting oscillation near 1 THz dominates the subsequent $\tau$. The oscillation at $\omega_{\rm pump}= 0.5 \unit{THz}$ may indicate a broken inversion symmetry, which warrants further investigation. The long-lasting oscillation near 1 THz could be either 2$\omega_{\rm pump}$ or 2$\omega_{\rm JPR}$, which will be further testified by pump-wavelength and temperature-dependent experiments below. Figure \ref{Fig:3} (g) shows the two-dimensional $\Delta E(\omega_t, \omega_{\tau})$. The bright spots at (0.5, $-$1) and ($-$0.5, 1) are nearly round and not stretched, which indicates the oscillatory central frequency along $\tau$ axis is almost a constant as shown in Fig. \ref{Fig:3} (d). The inset of Fig. \ref{Fig:3} (g) shows the calculation result of $\equa{eqn:chi3}$, with $\omega_{\rm pu} = 0.5$ THz, $\omega_{\rm pr} = 1$ THz, $W_{\rm pu} = 0.1$ THz, and $W_{\rm pr} = 1$ THz, whose pattern is qualitatively the same with Fig. \ref{Fig:3} (g). According to $\equa{eqn:chi3}$, the limited bandwidth of pump pulses results in the constant oscillatory central frequency.

\begin{figure}[t]
  \centering
\includegraphics[width=8cm, trim=0 0 0 0,clip]{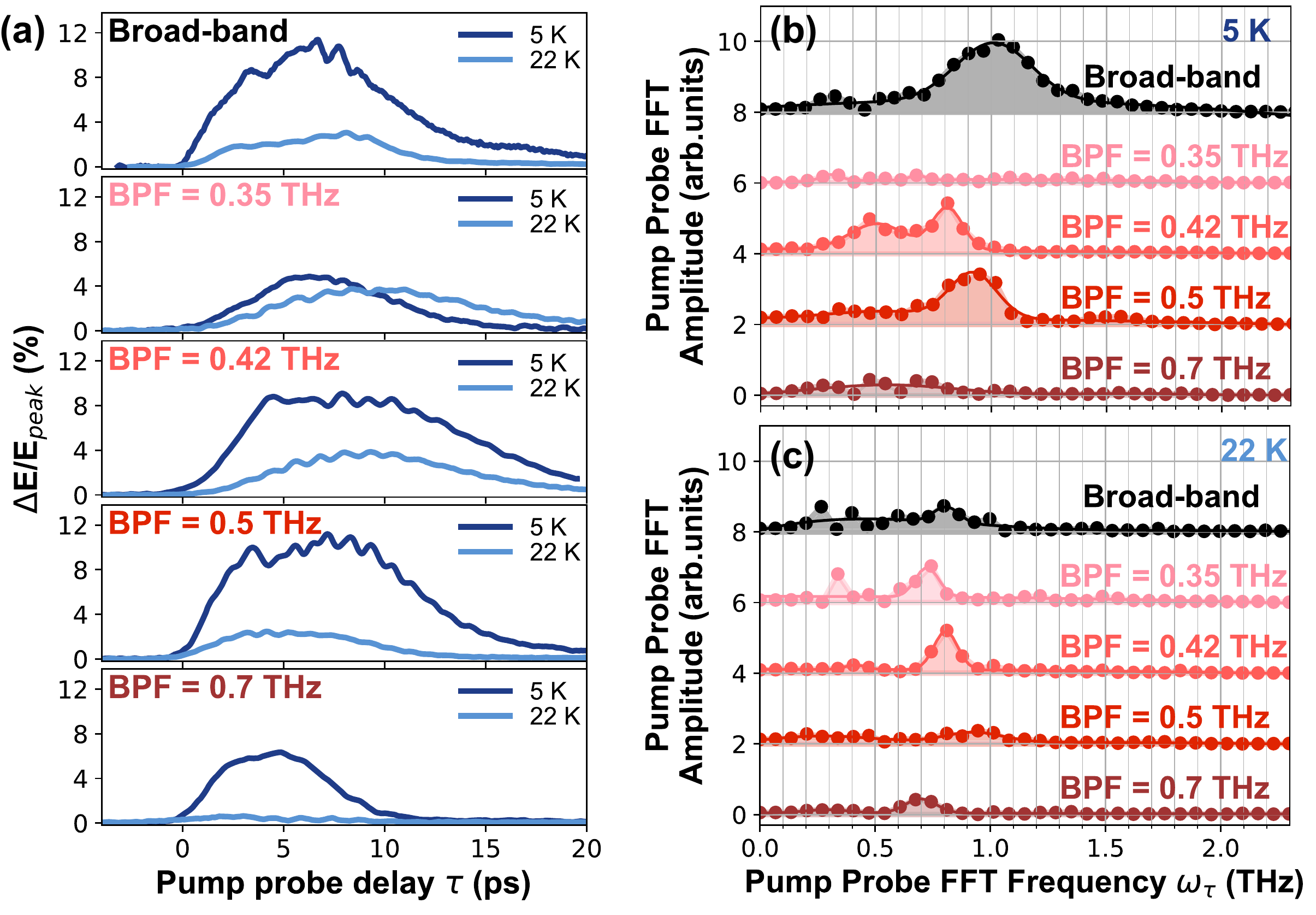}\\
\setlength{\abovecaptionskip}{0.1cm}
  \caption{Pump-wavelength and temperature-dependent experiments. (a) The pump-probe decay profiles of $\Delta E(t = 4.2$ ps, $\tau)$ after different excitations at 5 and 22 K. FFT amplitude of the extracted oscillatory signals in $\Delta E(t = 4.2$ ps, $\tau)$ at (b) 5 K and (c) 22 K. }\label{Fig:4}
\end{figure}

Figure \ref{Fig:4} (a) shows the pump-probe decay profiles of $\Delta E(t = 4.2$ ps, $\tau)$ at 5 and 22 K after the excitations of different THz pump, which is obtained by fixing $t$ to 4.2 ps and scanning over $\tau$. Figure \ref{Fig:4} (b) and (c) present the amplitude of oscillatory signals extracted from Fig. \ref{Fig:4} (a). For comparison, a temperature-dependent measurement was firstly performed for the single-cycle THz pump case, in which the oscillation central frequency shifts from 1 to 0.8 THz due to a redshift of JPM as temperature increases from 5 to 22 K.

For the multi-cycle THz pumps centered at 0.35 and 0.7 THz, which are significantly off-resonant with $\omega_{\rm JPR}$ at 5 K, nearly no oscillatory signals can be recognized at 5 K. Upon tuning $\omega_{\rm pump}$ to 0.42 THz, weak oscillations centering at 0.8 THz show up. For 0.5-THz pump resonant with $\omega_{\rm JPR}$ at 5 K, the oscillations is maximized, with a central frequency of 1 THz. The pump wavelength-dependent experiment confirms that the long-lasting oscillations  induced by multi-cycle THz pump do peak around 2$\omega_{\rm pump}$. Upon increasing temperature to 22 K, JPM shifts to $\sim$ 0.4 THz and becomes more damped. The oscillation amplitude at 22 K reaches a maximum when $\omega_{\rm pump}$ = 0.42 THz. For the 0.35-THz pump, clear oscillations at 0.7 THz emerge, quite different from the non-oscillating behaviour at 5 K. For the 0.5-THz pump, oscillations still peak at 1 THz but get almost invisible. As the temperature approaches T$_c$, JPM moves out of the measurement range and nearly no pump-induced changes are detectable. To summarize, for multi-cycle THz pump centering at $\omega_{\rm pump}$, the central frequency of $\tau$-axis long-lasting oscillations is equal to 2$\omega_{\rm pump}$. The oscillation amplitude is maximized when $\omega_{\rm pump}$ resonates with $\omega_{\rm JPR}$, which could result from the filtering effect imposed by JPE and the emission coefficient $F(\omega, \mathbf{k})$.

We now discuss the origin of the long-lasting ($\sim 20 \unit{ps}$) temporal oscillations in the pump-probe process which is also revealed by the sharp lines in Fig.~2h with widths $\sim 0.1 \unit{THz}$. We proposed that the  oscillations are due to the sharp peaks in the emission coefficient $F$ in \equa{eqn:chi3_time}.  Another possible reason is the transmission peak (Fig. \ref{Fig:1} (d)) of the interface around JPM which may lead to a narrow frequency signal inside the sample even for a broadband single-cycle pump. However, the width of the transmission peak is far too broad to explain it.

Alternatively, the oscillations may come from  excitation of a long lived collective mode around $\omega_{\text{JPR}}$. Of course, the EM waves inside the sample excited by the pump and probe  may be called (hyperbolic) Josephson plasmons\cite{Sun.2020superconductor}, but they span a wide frequency range above $\omega_{\text{JPR}}$ instead of being a single mode (SI Sec.~3.4.1). Inhomogeneity in the sample caused by disorder may introduce some longitudinal components to the incident field to excite the surface JPMs (SI Sec.~3.4.2). However, if these effects existed, they should have added a Lorentzian oscillator around $\omega_{\text{JPR}}$ to the optical conductivity and corresponding features to the linear reflectivity~\cite{cremin2019photoenhanced}, which are absent in \fig{Fig:1}(d).
In the nonlinear pathway, JPMs with opposite momenta may be parametrically excited in pairs by the pump~\cite{gabriele2021non, Michael.2020, Dolgirev:2022wo}. Again, since the JPMs have a wide dispersion covering a large frequency range, this process would result in a continuous exciting spectra instead of a well-defined resonance around $\omega_{\text{JPR}}$. Moreover, being proportional to quantum fluctuations of the JPM, the plasmon-pair contribution~\cite{gabriele2021non} is much weaker than the tree level non-linearity considered here (SI Sec. 3.3). The relevance of this interesting contribution \cite{gabriele2021non} in our experiment requires future investigation. Finally, the Higgs amplitude mode has been excited nonlinearly before using strong THz radiation in conventional superconductors~\cite{Higgs_Shimano_single,Higgs_NbN_science,Higgs_Nb3Sn_JGWang,PhysRevB.104.L140505,PhysRevB.105.L100508,PhysRevLett.109.187002} and within the CuO$_2$ layers of HTSCs~\cite{PhysRevLett.120.117001,chu2020phase,chu2021fano}. Unfortunately, Higgs mode is irrelevant here since its frequency $\sim 1.7 \Delta_{0}$~\cite{PhysRevB.102.014511} is much higher than those of the THz pump pulses in our experiments  where 2$\Delta_0$ = 30 meV is the gap at the anti-nodal position \cite{valla2006ground}. This conclusion is further reinforced by the multi-cycle THz pump experiment, which demonstrates a resonance in the oscillation when $\omega_{\rm pump}$ approaches $\omega_{\rm JPR}$.

In conclusion, in the THz pump-THz probe experiments performed on La$_{1.905}$Ba$_{0.095}$CuO$_4$, long-lasting temporal oscillations are observed in the out-of-plane transient responses of the superconducting state during the pump-probe decay process, which results from the c-axis third order nonlinear response together with interesting frequency dependence of the emission coefficient. For the single-cycle THz pump, the oscillation frequency is linearly dependent on the frequency-of-detection of the THz probe beam and also closely related to the frequency of the JPE, rather than centering at a fixed frequency. In contrast, the multi-cycle THz pump centering at $\omega_{\rm pump}$ leads to the most dramatic oscillations once $\omega_{\rm pump}$ approaches $\omega_{\rm JPR}$. We conclude that the observed oscillations in the pump-probe delay probably come from the sharp emission peak close to the JPM, an EM phenomenon, while the resonant excitation of a single collective mode is not necessary to explain it.  The result of this study not only provides a benchmark of nonlinear optical responses of HTSCs but also offers new insight into the emerging frontier of strong-field THz spectroscopy on related complex quantum materials. In the future, with \equa{eqn:chi3} applied to single color THz pump-probe spectroscopy, it will be possible to map out the nonlinear susceptibility quantitatively, offering more direct information of the material.

\section{Methods}
The strong-field THz pump with stable carrier-envelope phase is generated by tilted pulse-front method. The THz probe was generated and detected by ZnTe crystals. The polarization of the THz pump and probe is set to be parallel to the $c$-axis of the crystal. The detailed schematic optical path diagram can be found in SI. The single crystal is grown using the traveling-solvent floating-zone method. The c-axis is acquired by cutting the a-b surface at 90$^\circ$ with a subsequent polishing.

\section{Supplementary information}
Supplementary information is available at NSR online. 

\section{Acknowledgments}
We would like to thank Z. L. Li and Y. Wan for fruitful discussions. This work was supported by the National Natural Science Foundation of China (No. 11888101), the National Key Research and the Development Program of China (No. 2017YFA0302904). The work at BNL was supported by the US Department of Energy, Office of Basic Energy Sciences, contract no. DOE-sc0012704. S. J. Zhang was also supported by China Postdoctoral Science Foundation (No. 2020M680181). The theoretical work at Tsinghua University was supported by the startup grant from the State Key Laboratory of Low-Dimensional Quantum Physics and Tsinghua University.

\section{Author contributions}
S.J.Z., T.D., and N.L.W designed research; G.D.Gu grew the single crystal; S.J.Z and Q.M.L built the experimental setup; S.J.Z performed experimental measurements with the help of Q.M.L, Z.X.W, Q.W, L.Y, S.X.X, T.C.H, R.S.L, X.Y.Z, J.Y.Y and T.D; Z.S. performed the theoretical calculations; S.J.Z. and Z.S. analyzed data; N.L.W provided overall guidance on the project; S.J.Z., Z.S. and N.L.W. wrote the paper with input from all authors.

\bibliography{terahertz}

\end{document}